\newcommand{\CLASSINPUTtoptextmargin}{1in}
\newcommand{\CLASSINPUTbottomtextmargin}{1in}
\newcommand{\CLASSINPUTinnersidemargin}{1in}
\newcommand{\CLASSINPUToutersidemargin}{1in}

\documentclass[10pt, conference, compsocconf]{IEEEtran}

\ifCLASSINFOpdf
  \usepackage[pdftex]{graphicx}
  \graphicspath{{./images/}}
  \DeclareGraphicsExtensions{.pdf,.jpeg,.png,.jpg}
\else
  \usepackage[dvips]{graphicx}
  \graphicspath{{./images/}}
  \DeclareGraphicsExtensions{.eps}
\fi

\begin{document}

\title{Usage Management of Personal Health Records}

\author{\IEEEauthorblockN{Christopher C. Lamb, Gregory L. Heileman}
\IEEEauthorblockA{University of New Mexico\\
Department of Electrical and Computer Engineering\\
Albuquerque, NM 87131-0001 \\
\{cclamb, heileman\}@ece.unm.edu}
\and
\IEEEauthorblockN{Pramod A. Jamkhedkar}
\IEEEauthorblockA{Princeton University\\
Department of Electrical Engineering\\
Princeton, New Jersey 08544 \\
pjamkhed@princeton.edu}
}

\maketitle

\begin{abstract}
Personal health record (PHR) management is under new scrutiny as private companies move into the market and government agencies actively address perceived health care distribution inequalities and inefficiencies.  Current systems are coarse-grained and provide consumers very little actual control over their data.  Herein, we propose an alternative system for managing the use of healthcare information.  This novel system is finer grained, allows for data mining and repackaging, and gives users more control over their data, allowing it to be distributed to their specifications.  In this paper, we outline the characteristics of such a system in different contexts, present relevant background information and research leading to the system design, and cover specific usage scenarios supported by this system that are difficult to control using simpler access control strategies.
\end{abstract}


\section{Introduction}
New healthcare legislation has spurred previously unknown levels of public and private investment in technologies supporting more efficient healthcare delivery \cite{Emr:Web:Recovery}.   An active area of examination is personal health records (PHRs).  Current systems, such as Microsoft HealthVault and the late Google Health are and were a start in this area, but provide rudimentary control over health information, provide consumers with very little actual control of their information, and essentially demand proprietary lockin to these products because of the amount of effort involved with data transfer \cite{Emr:EvaluationHealthInf}.

In this paper we describe a new framework that allows for an open, consumer-centric approach to health information storage and consumption centered around flexible and fine-grained usage management policies, as well as an implementation of that framework.  User empowering systems in this area are needed to allow users control over the information that represents them, and would be in high demand if appropriately designed \cite{Emr:PyAmWaCr}.  We propose to address this need by bundling health information (either entire records or subsets of records) with traceable and aggregateable usage policies controlled by the users themselves.  Users would have the ability to make aspects of their records available to everyone from research institutions looking for historical information for studies, to healthcare providers who need specific information to support diagnoses.  Such a framework will enable the combination of information from groups of users and determine dynamically via policy evaluation how that new set of data can be used in a way that complies with all included user policies.  If the combined dataset cannot be used, policies can be analyzed to determine the cause of policy conflict.

The \textbf{primary research contribution of this work} is both the \textit{application} of usage management principles and the \textit{demonstration} of the feasibility of realistically applying these principles in a unified system architecture.  Our work establishes the importance of fine grained usage management in the medical domain, demonstrates the advantages of systems that can manage PHRs in this way, describes in a general sense how this can be accomplished, and finally reviews an operational system demonstrating these capabilities.

Herein, we describe the design of a novel system that supports fine-grained management of those data elements in a PHR, and we demonstrate how this can be used in two example applications.  The proposed system will allow users to specify policies over the data itself rather than the entire record in question, providing control over information dissemination.  We will demonstrate the above mentioned capabilities in two distinct usage scenarios.  The first includes two distinct parties negotiating over access to specific information contained in a health record.  If the parties reach an agreement, the information consumer is granted access to specific medical data, for an agreed-upon price.  The second demonstrates a data broker combining a set of previously acquired health record data into an aggregate set for research, if the licensure is in fact compliant between all selected data elements.  In the second usage scenario, we will discuss the constraints associated with inserting aggregate data sets derived from PHRs back into a marketplace.

This kind of system, allowing users control over their data in ways fostering ease of dissemination, use and reuse, helps users receive better, more targeted care, helps providers easily access required information, and allows this kind of data to be more easily examined and mined.  We apply established system design principles, used in the development of Internet-scale networks to create a open flexible system \cite{Al:04,BlCl:01,ClWrSoBr:02}.  We standardize certain features, such as operational semantics and logical data domains, but otherwise limit the impact of the policy system on data dissemination as much as possible.

\subsection{Previous Work}
Even though automated usage management of PHRs is an evolving concept, there exists a rich body of previous literature in the fields of usage management, DRM, and access control. The proposed framework leverages the results in these areas to address the specific challenges posed in usage management of PHRs. Most of the research applicable to the combination of previous artifacts into a single aggregate artifact comes from the DRM world in particular.  Generally, these expressive languages have been fundamentally based on different types of mathematical logic or formalisms with reasoning capabilities \cite{ArHu:07,BaMi:06,ChCoEtHaJoLa:03,HaWe:04,XiBjFu:08}.  This approach, while useful in closed systems, tends to not work as usefully in more open dynamic environments.  This has led to the development of translation mechanisms to address interoperability needs \cite{HeJa:05,PoPrDe:04,ScTaWo:04}.  This translation process is difficult for most policy languages, and in fact infeasible as a result \cite{KoLaMaMi:04,SaShUe:04}.  Alternative approaches have required the use of sophisticated and powerful languages that must be adopted as a universal standard \cite{OMADRM,ODRL-req,Wa:04,XrML-spec}.  This approach inherently limits innovation and flexibility \cite{HeJa:05,JaHe:04,JaHe:08,JaHeMa:06}.  Usage Management approaches eclipse standard access control methodologies for use in this type of domain, as access control is a necessary but not sufficient technology supporting this kind of asset management \cite{PaSa:04,BL:73,BL:76}.  Recent work applying DRM concepts to healthcare records with respect to encryption-centric content protection and partitioning is in fact complementary and serves to validate concepts contained herein \cite{Jafari:2011:RMA:2046631.2046637}.

\section{New Models}
Engineers and futurists have speculated as to the impact of PHRs for years \cite{Emr:Web:BestCaseEMR,Emr:Web:WorstCaseEMR}.  Others have speculated on the institutional use of PHRs by organizations in today's regulated medical environment \cite{Emr:doi:10.1056/NEJMc081118}.  Health records, when under the control of the person they address, are no longer controlled by the Health Insurance Portability and Accountability Act (HIPPA), though the companies that manage them on the user's behalf in these cases are regulated in most aspects by the Electronic Communications Privacy Act \cite{Emr:doi:10.1056/NEJMsb0800220}.  In total, these concerns imply certain requirements on robust health record systems, making usage models and record control more complex.  None of the promises or concerns of PHRs can be realized or mitigated without strong usage management.  With a dependable usage management capability, PHRs open new horizons in the services landscape for interested adopters.

\subsection{A Note on Reliability}
In order for PHRs to be effective, they must be actively used by health care providers.  A system with the wrong kinds of editability constraints or auditing capabilities is at risk of remaining unused by an individual's care providers.  Ideally, these kinds of health records would contain the kind of information a physician would include in a patient's chart.  Providers are required to maintain this is information for adequate patient treatment.  If this information can be arbitrarily edited however, it loses its credibility.

In fact, many employer-sponsored monitoring programs may incentivize gold-plating medical histories.  Systems like Virgin HealthMiles are marketing themselves directly to employers as ways to monitor employee health \cite{Emr:Web:VirginHealthMiles}.  Companies are using Virgin HealthMiles to track employee exercise, and as an incentive to use the product (and get more exercise), are offering additional contributions to employer-sponsored health savings accounts if employees meet certain criteria.  Similar scenarios could be right around the corner for personal health management systems, were employers incentivize employees to decrease blood pressure, change diet, or similar kinds of things.  In those situations, the pressure for users to alter their records to reflect the reality their employers want to see will be significant, and many users are likely to resort to embellishing their records as a result. Once that happens, health care providers can no longer use the records to provide care.

Any system managing these kinds of records must therefore provide mechanisms to certify, if not the accuracy of the provided information, at least the veracity of it.  Care providers must be able to trust the information provided in a given record, and must not be required to shoulder the burden of viewing the record's edit history in order to do so.  This implies a separation of roles between those who can edit the content of a given record, and those who control how the content of that record may be used.

\subsection{Remote Information Access}
Remote access to a patient's health care information is a standard feature of everyday life to which most of us pay little attention.  While in school, we are required to provide evidence of vaccination.  When older, travel to most parts of the world requires rounds of injections.  Most travelers are strongly advised to purchase additional travel insurance to ensure appropriate care in emergencies.  Certainly, when traveling to some parts of the world internet access can be difficult to acquire, but nevertheless such access is much more common now than it was even two years ago, and is becoming easier and easier to find with the proliferation of cellular telephone networks in heretofore undeveloped countries.  

Open access to this kind of healthcare information would certainly make these scenarios easier to deal with for any user, but require strong usage management protections to be effective.  In each case we have distinct sets of users that require access to care information, and in each case those users require access to a specific and limited sections of a personal healthcare record.  School administrators, for example, need to confirm the vaccination status of students.  This requires unfettered access to a student's vaccination history, but not to that student's psychiatric care or genetic record.  Likewise, travel visa providers may need access to similar information.  On the other hand, care providers no matter the country of origin require comprehensive care record access in order to provide timely and accurate care.  Furthermore, users have different requirements with respect to the speed of access.  School administrators have much less of an urgent, pressing need for care information that an Ethiopian doctor treating an injured patient.  

Importantly, access need not be granted permanently.  Both administrators and foreign care providers could be given general role-based access that can be removed when no longer necessary.

The ability to provide care information in a secure, manageable way in these scenarios saves users significant time and headache.  Rounding up and delivering vaccination records to school administrators is time consuming and stressful.  Receiving emergency health care in foreign countries is more than a little frightening.  Systems that can help ameliorate these kinds of situations would certainly be useful.  Furthermore, without controls over specific data elements composing a given record, these users cannot be appropriately limited in their access.

\subsection{Monitoring}
To constrain health care costs, some employers are beginning to implement holistic preventative health programs.  These programs are structured to attempt to lower overall healthcare costs for a large group of employees through regular screenings, exercise programs, and key health marker monitoring.  Employers are interested in monitoring indicators like triglyceride levels, serum cholesterol, HDL/LDL ratios, blood glucose, blood pressure, and the like.  Employee participation is not necessarily mandated, but can be encouraged through additional contributions to health savings accounts for participating employees.  In these cases, employers have specific things in which they have an interest. Employees on the other hand likely have information in their care records they very much want to keep out of their employers hands.  An employee, for example, may very much want the additional HSA contribution for her family, but is not inclined to let her employer know about her anti-depression medications or her recent treatment for alcohol dependency.

A dependable usage management system supports this kind of partitioned use.  With appropriate controls, this information can be centralized and controlled by the record owner, who can create limited access for employers.  Furthermore, this kind of information can be aggregated by the user over a period of years, demonstrating a pattern of healthy behavior, and perhaps making that record owner more attractive to future employers.  Sensitive information can still be controlled by limiting access.

\subsection{Custom Care}
When users have aggregated medical information at a single location, companies can now access information ranging from previous drug reactions, current prescription status, and the current state of any disease markers.  Users can also make any reactions to medications known essentially immediately, rather than having to wait until able to visit a particular physician.  This kind of information could allow pharmaceutical companies to offer innovative services tailoring medications specifically for individual needs.  For example, large doses of Niacin are common treatment for patients with high serum cholesterol characterized by low HDL and high LDL.  One of the unfortunate side effects of Niacin is facial flushing \cite{Emr:Web:Niacin}.  This is a common but by no means universal side effect of Niacin use.  It can be alleviated in many cases by ingesting Aspirin 20-30 minutes prior to taking niacin \cite{Emr:Web:Niacin}.  Patients that suffer from this particular side effect could have their dose customized and time-released to relieve this discomfort, but only if pharmaceutical providers know this is a patient's problem.

This kind of system promises to provide more responsive care at a potentially lower cost to patients, as long as patient data is available to custom medication providers.  A usage management scheme controlling specific information with respect to pharmaceutical providers could allow those providers to create custom medications for clients  better suited to those specific users.

\subsection{Data Marketplace}
The system we describe in the following sections incorporates a market to allow users and brokers to profit from the use of personal health data released under mutually acceptable terms, where usage policies accompany filtered data for either dynamic or static evaluation.  Usage policies themselves are essentially unlimited in how they describe the use of a specific health record.

\section{System Architecture}
Whatever the final technical attributes of this kind of system are, that system must at least embody certain attributes and requirements.  That architecture must furthermore facilitate usage management as a first-order attribute.

\subsection{Requirements and Attributes}
This type of system has a group of attributes and requirements that it must embody in order to fulfill the use scenarios outlined in the previous section.  Without these, the system would either be poorly received by the targeted user population or essentially non-functional.  Of course, simply aligning with these features does not guarantee widespread use; they are necessary, but not sufficient to guarantee widespread adoption.  Core features include:

\begin{itemize}

\item \textit{Editability} When considering a specific health record, certain fields of that record should be editable by the owner.  Other fields must only be editable by specific medical providers.  The data owner has complete control over who those providers are, as well as what roles they fulfill.  Certain predefined roles however have access to fields of the health record that the owner does not.  For example, a physician in the provider role is able to add information to a health record, and edit information that provider has previously entered.  The record owner cannot edit or change that information, but can edit personal contact information.

\item \textit{Roles} The system must have clearly partitioned roles related to ownership of specific areas of a given record.  These roles must be verifiable as well, either via uploading scanned credentials, indexing into professional registries, or direct provider contact.  For example, someone assigned to a provider role must be verified as a healthcare provider to prevent record owners from creating spoofed accounts to circumvent record protections.

\item \textit{Auditability} The system must be able to keep a clear record of who edited what, what those specific changes were, how they were made, and when.  This audit trail helps to establish the trustworthiness of the system as a whole, and can provide configuration management functionality over records.

\item \textit{Security} The system must take advantage of modern security systems as much as possible to provide additional control over assets.  Without providing adequate security, sensitive information could leak to unscrupulous third parties.  For example, a data owner that has been treated for sexually transmitted diseases in the past five years, but has in fact been married for a decade, may want to keep that information as sequestered as possible.

\item \textit{Accessibility} Some of the outlined scenarios directly imply wide accessibility geographically, while others require access to medical information from devices with a variety of form factors ranging from modern smartphones to tables to desktop computers.  Still others require data to be delivered through data-centric rather than human readable means.  Not providing this kind of open, widely accessible system makes data entry and eventual use prohibitively difficult.

\item \textit{Performance} Core functionality must be high performance.  Data entry must be engaging and responsive, or providers will simply abandon the system for other, more traditional ways of recording user information.  Data retrieval must be likewise snappy, especially in critical care scenarios.

\item \textit{Flexibility} This system and the data it manages can be used in a wide variety of contexts.  In order to remain relevant, it must be flexible enough to be easily integrated with other arbitrary systems in both de-facto standardized (e.g. RESTful access) and officially standardized (e.g. WS-* SOAP-centric access) ways.

\item \textit{Extensibility} It must provide programmatic interfaces to allow integrations with other currently unknown systems.  Not doing so limits how users can access and exploit the information they own.

\end{itemize}

The system as a whole provides management of PHRs, which must obviously be addressed by any proposed system architecture.  This management includes record creation, read, and update. It should not delete any data healthcare data, but other less sensitive information should be deletable.

Other features like notifications on certain predefined events or predefined integrations with systems like Twitter or Facebook may certainly be useful, but are not core functionalities the system must provide.  They should be able to be addressed via core extensibility mechanisms and programming interfaces however.  

\begin{figure}[!t]
\centering
\includegraphics[width=3in]{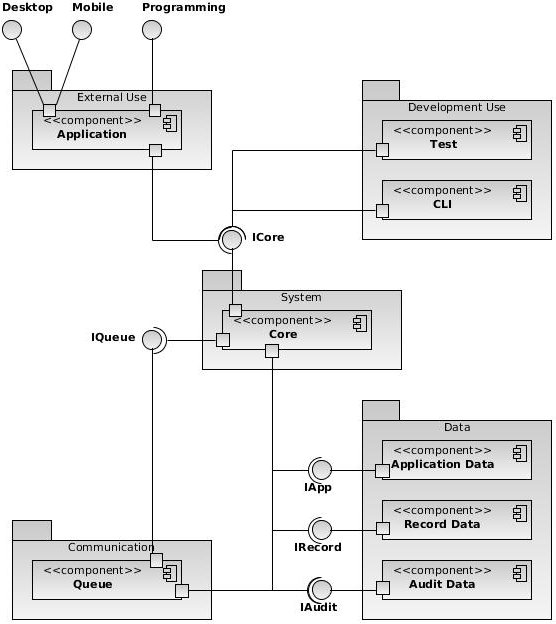}
\caption{System Architecture Runtime Component View}
\label{fig:RuntimeView}
\end{figure}

\subsection{Sample System Architecture}
Figure ~\ref{fig:RuntimeView} outlines a potential system architecture fulfilling the currently known attributes and requirements.  Our current system is in fact implemented against this specific system architecture, but implementations are by no means restricted to using our chosen toolset (i.e. the Ruby ecosystem) for implementation.  This runtime component view contains potential packages, interfaces, and components comprising a health record usage management system.  Note that this system architecture is currently technology agnostic.  Specific packages include:

\begin{itemize}

\item \textit{External Use} This package contains all user facing elements.  Mobile, desktop, and programming interfaces accessed by external users are contained in this package, as are all components providing this kind of functionality.

\item \textit{Development Use} This is another package containing system user access elements, but is intended to contain all private components used to access the system.  This package includes command line interfaces, allowing access via the interactive ruby shell, the Clojure read-eval-print-loop, or similar systems.  All system tests are defined here as well.  This includes unit tests, functional tests, and integration tests as well.

\item \textit{System} The system package contains the core system interface(s) and components.  These contain all the system specific logic, including any downloaded policies, licenses and the like.

\item \textit{Communication} Communication libraries and components are contained in this package.  Currently, we foresee the need for some kind of queuing component providing asynchronous data transfer between the core components and data stores.

\item \textit{Data} The final package contains data stores for the system.  We are using separate data stores to both enhance security and provide additional data storage flexibility.

\end{itemize}

Various components could be mapped to disparate technologies, even within the same package.  For example, the core system component could consist of a variety of ruby gems, running in an application hosted on Heroku, with tests in RSpec, and a command line interface to support granular user story experimentation.  Queuing components could be hosed on Amazon's Simple Queue Service (SQS), while Application data is hosted in a database in Heroku, Record Data via Cloud Files on Rackspace, and Audit Data in a store hosted via Amazon Simple Storage Service (S3).  The entire system would use standard security and trusted computing technologies.

\begin{figure}[!t]
\centering
\includegraphics[width=3in]{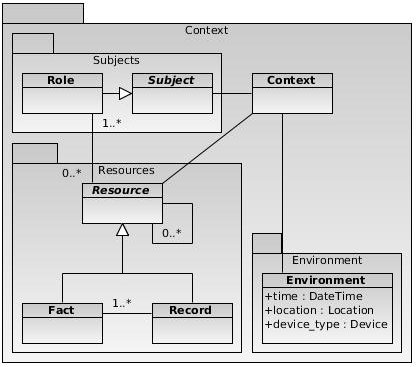}
\caption{System Usage Management Ontology}
\label{fig:Ontology}
\end{figure}

\subsection{Usage Management}
In order to apply usage management to a health record, we first define a logical data model over the records we plan to manage.  In Figure ~\ref{fig:Ontology}, we have outlined a static entity view for this domain.  In essence, we associate a group of \textit{facts} with a single \textit{record}, both of which are instances of managed \textit{resources}.  A \textit{resource} is associated with a \textit{role}, and that is an instance of a \textit{subject}.  Finally, all \textit{resources} and \textit{subjects} are related to a \textit{context}, which also refers to a given \textit{environment}. As shown in Figure~\ref{fig:Ontology}, the Environment, Subject, and the Role, each carries a set of parameters whose values determine the manner in which resources may be used within the system. 

As a simple example, consider an environment with parameters, \textit{Date, Location,} and \textit{DeviceType}, and the roles that include \textit{patient}  and \textit{physician}. Given these Environment and Role parameters, it is possible to express usage rules based on constraints on these parameters. For example, a particular PHR may allow edits to a patient information to be carried out only by physicians from a certified device physically located in a hospital for a given period of time. 

Whereas this is an overly simplified example of usage rules, usage management of PHRs present a unique set of the usage rules specification, interpretation, enforcement, and reasoning requirements. One of the primary requirements is that users need to be able to express usage rules in a fine grained manner that accurately express their concerns regarding the  usage of PHRs. To this effect, there exist multiple languages that allow expression of usage control rules on data sets~\cite{XrML-spec,PaSa:04,JaHeLa:10}.  PHRs are generally managed and used in a variety of environments that may not be known to a user \textit{a priori}, at the time of specification of the rules. Therefore, it is necessary that the usage rules can be interpreted in different types of environments, and they remain with the PHR throughout their lifecycle. Finally, PHRs are often managed in groups for the sake of research studies, and across hospitals, in such situations, it is necessary to interpret the common policies for a group of PHRs based on the usage rules associated with individual PHRs. In order in incorporate such dynamics, it is necessary that the language used for specification of usage rules is capable of handling data transformations such as mashups~\cite{JaHe:08}. 
 
\subsection{System Architecture and Attributes}
The system architecture described fulfills the various driving requirements and attributes.  First, we support finely managed \textit{editability} constraints via the association of \textit{facts} with \textit{roles}.  \textit{Roles}, as shown in Figure ~\ref{fig:Ontology}, are first-order system entities.  The system architecture supports \textit{auditability} via the Queueing and Audit Data components and the IQueue and IAudit interfaces.  \textit{Security} emerges from the system architecture via data partitioning and the use of standard security and trusted computing technologies.  We address \textit{accessibility} constraints via the Desktop, Mobile, and Programming application interfaces.  The system can flexibly address \textit{performance} issues within the Core component behind the ICore interface.  Judicious use of queuing enhances \textit{performance} as well.  Finally, \textit{flexibility} and \textit{extensibility} emerge from the loosely coupled componentry and publicly exposed programming interfaces.

\section{Sample System - Data Marketplace}
A data marketplace for medical data through which users can generate revenue from exposing medical information of their choosing is one possible business model we previously described.  Here, as a proof of concept of our proposed system architecture, we incentivize PHR adoption via the use of a data marketplace.  We have three primary categories of roles in mind:

\begin{figure}[!t]
\centering
\includegraphics[width=2in]{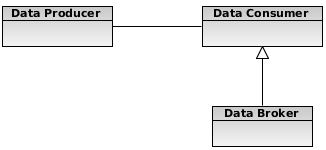}
\caption{Data Marketplace System Roles}
\label{fig:SysRoles}
\end{figure}

\begin{itemize}

\item \textit{Data Producers} who produce and market electronic medical information.  This category is generally limited expressly to individual users who require medical care and other related products.

\item \textit{Data Consumers} who directly consume medical information.  This category includes physicians, research institutions, and the like.

\item \textit{Data Brokers} who acquire and re-market medical data from data producers, making that data available in some kind of value-added way for other data consumers.  They are a proper subset of data consumers.

\end{itemize}

\textit{Data producers} would use the medical data market to profit from their personal information.  When negotiating over specifics concerning how their data can be used, they are free to manipulate any aspect of the usage terms prior to a final agreement with a \textit{data consumer}.  The \textit{data consumer} can accept or reject a specific proposal, as can a \textit{data producer}.  A typical negotiation would look something like this:

\begin{enumerate}

\item A \textit{data consumer} searches the marketplace for medical information meeting specific requirements.  This step is a call to a specific search interface in our example, but could be a manual process.

\item The search yields some results.  This proposed system returns a list of contact information of known \textit{data producers} that have data matching the search requirements.

\item The \textit{data consumer} initiates a negotiation for access to specific data.

\begin{enumerate}

\item The \textit{data consumer} contacts the \textit{data producer} and submits and initial proposal.

\item The \textit{data producer} responds to the initial proposal, either be indicating acceptance, rejecting the proposal, or submitting a counter proposal.

\item The \textit{data consumer} is then free to respond with acceptance, rejection, or a counterproposal of her own.

\end{enumerate}

\item Eventually, the negotiation will conclude with the parties having reached an agreement describing access to specific medical data with associated term or having failed to come to mutually acceptable terms with respect to data access.  

\end{enumerate}

Usage terms in a successful conclusion generally describe what the \textit{data consumer} can access, how they may use it and for how long, where it may be accessed, and so on.  It will also usually describe some kind of payment for use, which can be based on any arbitrary number of factors such as time, date, location, attribution, or perhaps in combination with other data.


\subsection{Implementation}
We have implemented this system and this specific case in the Ruby ecosystem.  The testcase itself is implemented in Cucumber, a case-oriented testing framework \cite{Emr:Web:Cucumber}.  The system itself has specific mappings from defined system components to technical implementations:

\begin{itemize}
\item \textit{Desktop, Mobile, Programming Interfaces} These interfaces are defined and implemented in the JavaScript Object Notation (JSON) using Representative State Transfer (REST) over HTTP as the communication protocol \cite{Emr:Web:JSON,Emr:Web:REST}.
\item \textit{External Use::Application Component} The application platform currently used is Ruby on Rails, though we are currently investigating migrating to Sinatra \cite{Emr:Web:RoR,Emr:Web:Sinatra}.
\item \textit{Development Use::Test Component} The test component has tests implemented in both Cucumber and RSpec \cite{Emr:Web:RSpec}.
\item \textit{Development Use::CLI Component} We use the Interactive Ruby Shell (IRB) for command line access if needed.
\item \textit{System::Core Component} Core components are hosted within our application server and distributed via the Ruby Gem standard \cite{Emr:Web:Gem}.
\item \textit{Communications::Queue Component} This is currently a ruby object.
\item \textit{Data Components} All data components are implemented in SQLite \cite{Emr:Web:SQLite}.
\end{itemize}

Other interfaces are use standard ruby semantics for access.  Databases in the Data package are in fact partitioned as shown in the system architecture model.

\subsection{System Data Model}
This system is built around a common data model that needs must be understood by any system developers.  It is currently used to define relationships and entities within the system at design and run time.  The \textit{record} elements in this ontology map to the \textit{record} elements in Figure ~\ref{fig:Ontology}, maintaining the one-to-many relationship of \textit{facts} to \textit{records}.  The primary elements in this ontology are:

\begin{itemize}

\item \textit{Producer} This is a data producer as defined in our user model.  A data producer owns a given \textit{record} that has been created over a lifetime of medical care.

\item \textit{Consumer} Again from the user model, a data consumer.  Data consumers use medical data in some way.

\item \textit{Record} A health record.  We can envision this as a set of discrete medical facts as defined in Figure ~\ref{fig:Ontology}.

\item \textit{Filter} A transformation of a health record.  If we have a record $ r $, we can transform $ r $ into $ r' $ by applying a transformation $ t $ such that $ r' = t(r) $ where $ t : record \rightarrow record $ and $ r' \subseteq r $.

\item \textit{Filtered Record} A filtered record is a record to which a filter has been applied.  If we have a filtered record $ r' $ derived from a record $ r $, then $ r' \subseteq r $.

\item \textit{License} A license describes the usage policy associated with a given filtered record.  This controls all aspects of filtered record use by an associated consumer.  The specific terms are negotiated over by the producer and the consumer until some optimal consensus is reached, and they then bind the use of an associated filtered record.  Licenses must provide the ability to trace use of transitively associated artifacts regardless of the degree of separation as well.  For example, if we have an artifact $ a $ composed of sets of data elements $ e_{0}, e_{1}, ... , e_{n} $ derived from records $ r_{0}, r_{1}, ... , r_{n} $, we need to be able to ensure that any use of a set of data elements $ e_{i}, i < n $ is within the policy bounds of record $ r_{i}, i < n $ and any compensation associated with such use is correctly attributed to the original data owners and brokers.

\item \textit{Bundle} A filtered record and associated license.  This is distributed to data consumers.

\end{itemize}

\begin{figure}[!t]
\centering
\includegraphics[width=3in]{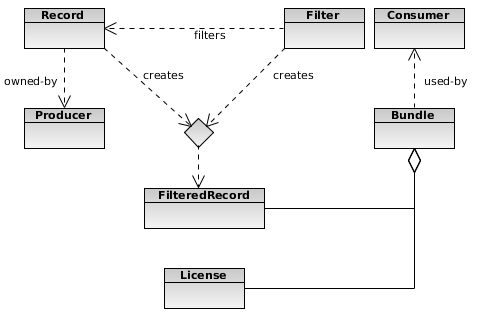}
\caption{Refined System Entity Model}
\label{System Ontology}
\end{figure}

\subsection{Dynamic and Static Policy Evaluation}
Usage policies can be evaluated over a spectrum bordered by two distinct approaches - either dynamically, at request time, or statically, when a bundle is created.  Pure dynamic policy evaluation evaluates the entire policy against an artifact at \textit{request time}, specifically and only when a request for an action is made by a consumer.  Static evaluation only occurs when \textit{the bundle is created} and is not evaluated at any later time.  While dynamic policies are more powerful, static policies are generally simpler to define, create, and apply.  Dynamic policy evaluation requires significant runtime infrastructure as well, which static evaluation will never require.  Furthermore, that runtime infrastructure must be present in a variety of systems, implemented upon a myriad of platforms in a slew of different programming languages.  Still, we have compelling reasons for developing dynamic evaluation systems.  Static systems cannot evaluate dynamic properties well.  Attributes like time are impossible to adjudicate with the simplest of static licenses and require some kind of dynamic evaluation.  Likewise, evaluation of a bundle's context is equally impossible to do with simple static policies.  Dynamic policies are more suitable for content that producers are interested in providing for unexpected use, while static policies generally only support predefined use scenarios.

In this system, we use a combination of static and dynamic approaches.  Static policy evaluation occurs immediately after negotiation between the producer and consumer, when a filter is applied to the health record.  This simplifies dynamic policy requirements by limiting the data that needs to be evaluated after the bundle is released.  If this filter were not applied, the dynamic policy would need to additional clauses to support hiding only those data elements to which the consumer has not been granted access.  All other evaluation occurs after the bundle is delivered to the consumer.  In order to support more complex and unexpected usage scenarios, including evaluating usage based on time constraints, this framework provides extensive dynamic evaluation capabilities after the initial filtering phase.  We also need to be able to support seamless operation over protected artifacts while disconnected from any kind of network or communication medium.  These factors lead to a powerful and local dynamic policy evaluation system.

\section{Conclusion}
New levels of heretofore unknown government interest in health care delivery will undoubtedly lead to increased use of personal health records \cite{Emr:Web:Recovery}.  Though current starts in this area are somewhat closed, limited, and suffer from a lack of user focus, organizations will develop new systems to deal with this extensive and valuable data.

In this paper, we have outlined access and editing concerns associated with personal health records as well as various new business models that proper usage management of personal health data can create.  These new models included remote monitoring, through which users have access to their medical information from remote locations over a variety of devices (including cellular telephones and the like).  We also described in some detail how both health care consumers and providers can benefit from long-term health monitoring, and finally covered how a PHR storage system incorporating usage management enables more targeted and precise medical care.  Without appropriate usage safeguards, these models would not be tenable.  We also covered in some detail the system architecture of a usage management system for health records and also described our current proof-of-concept implementation and the specific technologies used.

The system we have developed begins to address our outlined scenarios.  Here, we presented design elements of a specific implementation of our system, a data marketplace.  Our marketplace had specific roles associated with data provision and consumption, and incorporates agent-based negotiation over data access.

Future areas of study in this domain include extension of our current system to other outlined scenarios.

\bibliographystyle{abbrv}
\bibliography{emr,drm}

\begin{thebibliography}{10}
\providecommand{\url}[1]{#1}
\csname url@samestyle\endcsname
\providecommand{\newblock}{\relax}
\providecommand{\bibinfo}[2]{#2}
\providecommand{\BIBentrySTDinterwordspacing}{\spaceskip=0pt\relax}
\providecommand{\BIBentryALTinterwordstretchfactor}{4}
\providecommand{\BIBentryALTinterwordspacing}{\spaceskip=\fontdimen2\font plus
\BIBentryALTinterwordstretchfactor\fontdimen3\font minus
  \fontdimen4\font\relax}
\providecommand{\BIBforeignlanguage}[2]{{%
\expandafter\ifx\csname l@#1\endcsname\relax
\typeout{** WARNING: IEEEtran.bst: No hyphenation pattern has been}%
\typeout{** loaded for the language `#1'. Using the pattern for}%
\typeout{** the default language instead.}%
\else
\language=\csname l@#1\endcsname
\fi
#2}}
\providecommand{\BIBdecl}{\relax}
\BIBdecl

\bibitem{Emr:Web:Recovery}
{{F}ederal {G}overnment of the {U}nited {S}tates of {A}merica}, ``{T}racking
  the {M}oney,'' {http://www.recovery.gov}, December 2010.

\bibitem{Emr:EvaluationHealthInf}
A.~Sunyaev, A.~Kaletsch, and H.~Krcmar, ``{C}omparative {E}valuation of
  {G}oogle {H}ealth {API} vs. {M}icrosoft {H}ealthvault {API},'' in
  \emph{{P}roceedings of the {T}hird {I}nternational {C}onference on {H}ealth
  {I}nformatics}, ser. HealthInf 2010.\hskip 1em plus 0.5em minus 0.4em\relax
  Setubal, Portugal: INSTICC, 2010, pp. 195--201.

\bibitem{Emr:PyAmWaCr}
P.~C., J.~Amery, M.~Watson, and C.~Crook, ``{A}ccess to {E}lectronic {H}ealth
  {R}ecords in {P}rimary {C}are - {A} {S}urvey of {P}atients' {V}iews,''
  \emph{Medical Science Monitor}, vol.~10, no.~11, 2004.

\bibitem{Al:04}
H.~Alverstrand, ``The role of the standards process in shaping the internet,''
  \emph{Proceeding of the IEEE}, vol.~92, no.~9, pp. 1371--1374, 2004.

\bibitem{BlCl:01}
M.~S. Blumenthal and D.~D. Clark, ``Rethinking the design of the {I}nternet:
  {T}he end-to-end arguments vs. the brave new world,'' \emph{ACM Transactions
  on Internet Technology}, vol.~1, no.~1, pp. 70--109, Aug. 2001.

\bibitem{ClWrSoBr:02}
D.~D. Clark, J.~Wroclawski, K.~R. Sollins, and R.~Braden, ``Tussle in
  cyberspace: Defining tomorrow's internet,'' in \emph{{SIGCOMM}}, Pittsburg,
  Pennsylvania, USA, Aug. 2002, pp. 347--356.

\bibitem{ArHu:07}
A.~Arnab and A.~Hutchison, ``Persistent access control: A formal model for
  drm,'' in \emph{DRM '07: Proceedings of the 2007 ACM workshop on Digital
  Rights Management}.\hskip 1em plus 0.5em minus 0.4em\relax New York, NY, USA:
  ACM, 2007, pp. 41--53.

\bibitem{BaMi:06}
A.~Barth and J.~C. Mitchell, ``Managing digital rights using linear logic,'' in
  \emph{LICS '06: Proceedings of the 21st Annual IEEE Symposium on Logic in
  Computer Science}.\hskip 1em plus 0.5em minus 0.4em\relax Washington, DC,
  USA: IEEE Computer Society, 2006, pp. 127--136.

\bibitem{ChCoEtHaJoLa:03}
C.~N. Chong, R.~Corin, S.~Etalle, P.~Hartel, W.~Jonker, and Y.~W. Law,
  ``License{S}cript: {A} novel digital rights language and its semantics,'' in
  \emph{Third International Conference on the Web Delivery of Music}, Los
  Alamitos, CA, Sept. 2003, pp. 122--129.

\bibitem{HaWe:04}
J.~Y. Halpern and V.~Weissman, ``A formal foundation for {XrML} licenses,'' in
  \emph{Proceedings of the 17th IEEE Computer Security Foundations Workshop},
  Asilomar, CA, June 2004, pp. 251--265.

\bibitem{XiBjFu:08}
J.~Xiang, D.~Bjorner, and K.~Futatsugi, ``Formal digital license language with
  {OTS/CafeOBJ} method,'' in \emph{Proceedings of the sixth ACS/IEEE
  International Conference on Computer Systems and Applications}, Doha, Qatar,
  Apr. 2008.

\bibitem{HeJa:05}
G.~L. Heileman and P.~A. Jamkhedkar, ``{DRM} interoperability analysis from the
  perspective of a layered framework,'' in \emph{Proceedings of the Fifth ACM
  Workshop on Digital Rights Management}, Alexandria, VA, Nov. 2005, pp.
  17--26.

\bibitem{PoPrDe:04}
J.~Polo, J.~Prados, and J.~Delgado, ``Interoperability between {ODRL} and
  {MPEG-21} {REL},'' in \emph{Proceedings of the first international ODRL
  workshop}, Vienna, Austria, Apr. 2004.

\bibitem{ScTaWo:04}
A.~U. Schmidt, O.~Tafreschi, and R.~Wolf, ``Interoperability challenges for
  {DRM} systems,'' in \emph{IFIP/GI Workshop on Virtual Goods}, Ilmenau,
  Germany, 2004, http://virtualgoods.tu-ilmenau.de/2004/program.html.

\bibitem{KoLaMaMi:04}
R.~H. Koenen, J.~Lacy, M.~MacKay, and S.~Mitchell, ``The long march to
  interoperable digital rights management.'' \emph{Proceedings of the IEEE},
  vol.~92, no.~6, pp. 883--897, 2004.

\bibitem{SaShUe:04}
R.~Safavi-Naini, N.~P. Sheppard, and T.~Uehara, ``Import/export in digital
  rights management,'' in \emph{Proceedings of the Fourth ACM Workshop on
  Digital Rights Management}, Washington, DC, Oct. 2004, pp. 99--110.

\bibitem{OMADRM}
O.~M. Alliance, ``Enabler release definition for {DRM} {V}2.0,'' Open Mobile
  Alliance, Tech. Rep., April 2003,
  http://xml.coverpages.org/OMA-ERELD\_DRM-V2\_0\_0-20040401-D.pdf.

\bibitem{ODRL-req}
``Open digital rights language {ODRL} version 2 requirements,'' ODRL, Feb.
  2005, http://odrl.net/2.0/v2req.html.

\bibitem{Wa:04}
X.~Wang, ``{MPEG-21} rights expression language: Enabling interoperable digital
  rights management,'' \emph{IEEE Multimedia}, vol.~11, no.~4, pp. 84--87,
  Oct.\slash Dec. 2004.

\bibitem{XrML-spec}
``e{X}tensible {R}ights {M}arkup {L}anguage ({XrML}) 2.0 {S}pecification,''
  Contentguard, November 2001, http://www.xrml.org.

\bibitem{JaHe:04}
P.~A. Jamkhedkar and G.~L. Heileman, ``{DRM} as a layered system,'' in
  \emph{Proceedings of the Fourth ACM Workshop on Digital Rights Management},
  Washington, DC, Oct. 2004, pp. 11--21.

\bibitem{JaHe:08}
------, \emph{Handbook of Research on Secure Multimedia Distribution}.\hskip
  1em plus 0.5em minus 0.4em\relax {IGI} Publishing, 2008, ch. Rights
  Expression Languages.

\bibitem{JaHeMa:06}
P.~A. Jamkhedkar, G.~L. Heileman, and I.~Martinez-Ortiz, ``The problem with
  rights expression languages,'' in \emph{Proceedings of the Sixth ACM Workshop
  on Digital Rights Management}, Alexandria, VA, Nov. 2006, pp. 59--67.

\bibitem{PaSa:04}
J.~Park and R.~Sandhu, ``The {UCON}$_{ABC}$ usage control model,'' \emph{ACM
  Trans. Inf. Syst. Secur.}, vol.~7, no.~1, pp. 128--174, 2004.

\bibitem{BL:73}
D.~E. Bell and L.~J.~L. Padula, ``Secure computer systems: Mathematical
  foundations, {MTR-2547, Vol. I},'' The MITRE Corporation, Bedford, MA, Tech.
  Rep., March 1, 1973, eSD-TR-73-278-I.

\bibitem{BL:76}
------, ``Secure computer system: Unified exposition and multics
  interpretation, {MTR-2997, Rev. 1},'' The MITRE Corporation, Bedford, MA,
  Tech. Rep., March 1976, eSD-TR-75-306.

\bibitem{Emr:Web:BestCaseEMR}
J.~Powers, ``{G}oogle {H}ealth 2018: {B}est {C}ase {S}cenarios,''
  {http://in3.org/articles/gh2018best.htm}, May 2008.

\bibitem{Emr:Web:WorstCaseEMR}
------, ``{G}oogle {H}ealth 2018: {W}orst {C}ase {S}cenarios,''
  {http://in3.org/articles/gh2018worst.htm}, June 2008.

\bibitem{Emr:doi:10.1056/NEJMc081118}
\BIBentryALTinterwordspacing
``{S}hifts in {H}ealth {I}nformation,'' \emph{New England Journal of Medicine},
  vol. 359, no.~2, pp. 209--210, 2008. [Online]. Available:
  \url{http://www.nejm.org/doi/abs/10.1056/NEJMc081118}
\BIBentrySTDinterwordspacing

\bibitem{Emr:doi:10.1056/NEJMsb0800220}
\BIBentryALTinterwordspacing
K.~D. Mandl and I.~S. Kohane, ``{T}ectonic {S}hifts in the {H}ealth
  {I}nformation {E}conomy,'' \emph{New England Journal of Medicine}, vol. 358,
  no.~16, pp. 1732--1737, 2008. [Online]. Available:
  \url{http://www.nejm.org/doi/abs/10.1056/NEJMsb0800220}
\BIBentrySTDinterwordspacing

\bibitem{Emr:Web:VirginHealthMiles}
``{V}irgin {H}ealth{M}iles,'' {http://us.virginhealthmiles.com}, January 2011.

\bibitem{Emr:Web:Niacin}
``{P}ubmed {H}ealth - {N}iacin,''
  {http://www.ncbi.nlm.nih.gov/pubmedhealth/PMH0000700}, January 2011.

\bibitem{JaHeLa:10}
P.~A. Jamkhedkar, G.~L. Heileman, and C.~Lamb, ``An interoperable usage
  management framework,'' in \emph{Proceedings of the Tenth ACM Workshop on
  Digital Rights Management}, Chicago, Oct. 2010.

\bibitem{Emr:Web:Jade}
``{J}ava {A}gent {DE}velopment {F}ramework,'' {http://jade.tilab.com/}, January
  2011.

\bibitem{Emr:Web:Fipa}
``{T}he {F}oundation for {I}ntelligent {P}hysical {A}gents,''
  {http://www.fipa.org}, January 2011.

\end{thebibliography}

\end{document}